# Title page

**Title:**

T2 Radiomic Features Are More Sensitive Than Mean T2 for Cartilage Load Response: A Stress MRI Study

**Authors**

Teresa Lemainque, Philipp Schad, Simon Westfechtel, Marc von der Stück, Robert Siepmann, Axel Honné, Christiane Kuhl, Daniel Truhn, Sven Nebelung

**Institutions**

[1] Diagnostic and Interventional Radiology, Medical Faculty, RWTH Aachen University, Aachen, Germany

[2] Scientific Workshop of the University Hospital Aachen, Medical Faculty, RWTH Aachen University, Aachen, Germany

[3] Molecular and Cellular Anatomy, Medical Faculty, RWTH Aachen University, Aachen, Germany

**Corresponding author:**

Teresa Lemainque

University Hospital RWTH Aachen

Department of Diagnostic and Interventional Radiology

Pauwelsstr. 30

52074 Aachen

E-mail: tlemainque@ukaachen.de

Tel. +49 241 80 37843

## Abstract

**Objective:**

To assess response-to-loading in a human cadaveric knee joint model under different loading conditions before and after meniscectomy

**Design:**

In this prospective study, stress magnetic resonance imaging was performed using an MR-compatible loading device and quantitative T2 mapping in unloaded (UL), 0° neutrally loaded (LN) and 10° varus loaded (LV) condition before and after meniscectomy. Mean T2 values and four radiomic texture parameters were assessed within the cartilage of medial femur (MF) and medial tibia (MT) for all conditions.

**Results:**

Medial joint space width decreased from UL to LN to LV and after meniscectomy (all $p<0.05$). T2 values did not show any significant dependency on pressure or meniscectomy (all $p>0.05$). The radiomic parameter 'variance' could assess loading induced textural T2 changes in the MF (UL vs. LN: $p=0.042$; UL vs. LV: $p<0.001$; LN vs. LV: $p=0.022$), and, in part, in the MT (LN vs. LV: $p<0.013$). Meniscectomy did not significantly alter the T2 mean values or radiomic parameters, respectively.

**Conclusions:**

T2-based radiomic features were more sensitive to assess cartilage response to loading than T2-mapping alone.

**Keywords:**

Stress MRI; cartilage; meniscectomy; T2 mapping; radiomics

# Introduction

Osteoarthritis (OA), a chronic degenerative joint disease, constitutes a rising health burden, with a global prevalence of over 500 million cases in 2019 [1]. The knee joint is the most often affected one. Here, OA results in progressive loss of cartilage, that is often regarded the hallmark of OA progression, and leads to an altered mechanical load distribution and hence pain and joint instability. Yet, OA is not restricted to cartilage only, but regularly involves other structures such as menisci, ligaments, and bones.

MRI is well-suited for whole-joint and articular cartilage assessment in particular; however, it detects OA-mediated degenerative cartilage alterations only in the irreversible stage, i.e., when cartilage thickness is already reduced [2]. Moreover, MRI examinations are performed in a per se unphysiological condition, i.e., when the patient is lying in supine position without subjecting the joints to the body weight.

Quantitative mapping techniques in MRI, such as T2 or T1rho mapping, promise to detect cartilage damage when it is still reversible, i.e., not yet visible on the conventional, morphological MRI sequences [3]. Adding a T2 mapping sequence to a routine knee MRI protocol has shown to increase its diagnostic performance, where in the context of early OA an increased sensitivity for cartilage softening, fibrillation and superficial cartilage defect were notable [4]. Beyond T2 mapping, T2-based radiomics, i.e., texture analysis of cartilage T2 maps, has been employed to diagnose (early) OA and to predict its progression or future incidence [5].

Stress MRI uses MRI-compatible loading devices to address the problem of unphysiological loading conditions. Joint laxity and stability may be evaluated by imaging the joint under standardized loading, which has shown to be promising in the evaluation of the cruciate and collateral ligaments [6, 7, 8, 9], cartilage [10, 11], and menisci [12, 13]. As far as meniscus deficiencies and postoperative cartilage damage are concerned, T2 mapping (yet without additional loading) has been evaluated in human subjects to assess cartilage damage before and after partial meniscectomy. Murakami et al. found elevated cartilage T2 values six months after arthroscopic lateral meniscectomy as compared to baseline values before the operation [14]. Based on data of the Osteoarthritis Initiative, Neumann et al. found higher T2 relaxation times in meniscectomized patients, indicating greater cartilage deterioration as compared to matched controls without surgery. In a study comparing T2 values before and after running between post-meniscectomy knees and healthy volunteers, Lindner et al. found running to cause significantly lower T2 values for the post-meniscectomy group, but no for the healthy group, which they related to excessive mechanical load after surgery [15]. Shiomi et al did employ stress MRI, studying the effect of pressure application before and after medial meniscectomy on response-to-loading patterns in porcine knee cartilage, and found location-dependent alterations in T2 changes post meniscectomy as compared to the intact state [16]. Bridging the gap between animal and human joint models, we have previously proposed a pneumatic loading device for human cadaveric knee joints that is able to vary both the strength and angulation of the load application in the *in-situ* setting, and assessed the effect of axial, varus and valgus loading on cartilage thickness alterations [11].

Thus, the aim of this exploratory study was to use this highly standardized and versatile loading device to investigate the influences of different mechanical loading status and meniscus integrity on the response-to-loading patterns in human cadaveric knee joints by means of T2 mapping and T2-based radiomics, which was performed with and without loading and pre and post meniscectomy, respectively. We hypothesized that loading would reduce the T2 values, and that meniscectomy would alter this response.

# Methods

## Study design

The institutional review board approved this prospective, exploratory study on human cadaveric knee joint specimens (AZ-EK180/16). Written informed consent of the body donors was available at the time of study initiation. Six fresh-frozen and visually intact human cadaveric knee joints were obtained from the department of Molecular and Cellular Anatomy, RWTH Aachen University. Moderate-to-severe cartilage degeneration, apparent by a markedly reduced cartilage thickness on the MR images, and a specimen length of less than 28 cm were defined as exclusion criteria. Prior to preparation, specimens were left to thaw for 24 hours. **Figure 1(d)** gives an overview over the workflow. After specimen preparation, a first MRI series was performed in three loading configurations. Afterwards, the menisci were removed arthroscopically. Finally, a second MRI series repeating the three loading configurations was performed. Between imaging and arthroscopy sessions, specimens were refrigerated at 4 °C.

## Specimen preparation

Prior to MRI experiments and arthroscopy, the knee joints were prepared as previously described [11]. In brief, knee joint specimens were cut to a length of 30 cm. Soft tissues around the femoral and tibial bone were dissected over a length of 5 cm and the fibula was shortened accordingly (**Figure 1(b)**). The femoral and tibial bone ends were fixed in the bone pots of the loading device while the specimen was in full extension and neutrally aligned using polymethlmethacrylate (Technovit 3040, Heraeus Kulzer GmbH, Wehrheim, Germany) and an additional spike attached to the bottom of the bone pots that was driven into the cancellous bone before curing of the fixation mass.

## MR-compatible loading device

We used an MR-compatible loading device for human cadaveric knee joints that has been described previously (**Figure 1(a)**) [11]. Confectioned for human cadaveric knee joints with a length of 28-32 cm, it allows for variable compressive in-situ loading and variable axis alignment. In this study, three loading configurations were investigated, i.e., unloaded under neutral (i.e., straight) alignment (UL), 3 bar loading with a pressure of 3 bar (or 0.74 kN, respectively) under neutral alignment (LN), and 3 bar loading with 10° varus alignment (LV). For loading under neutral alignment, two 0°-angled bone adapters were used, while two 5°-angled bone adapters were used for loading under varus alignment. For the unloaded configuration, the press-point discs of the loading device were in loose contact with the specimen without any force application.

## MRI measurements

After positioning a prepared knee joint in the loading device, measurements were performed on a 3.0 T clinical MRI scanner (Achieva, Philips, Best, The Netherlands). To that purpose, two flexible ring coils (SENSE Flex-L, Philips, Best, The Netherlands) were positioned left and right to the knee joint for MR signal reception. The loading device was positioned centrally in the MR scanner bore. There were two imaging sessions per knee joint specimen, i.e., i.e., the intact state prior to meniscectomy (INT) and post meniscectomy (PM), during which the same order of measurements was observed: first, the UL configuration was acquired, followed by LN and LV. Following pressure application, an equilibration period of 5 minutes was obeyed before proceeding with the image acquisition. Per knee joint and configuration, proton density-weighted fat-saturated (PDw fs) sequences for morphological assessment as per clinical reference standard and multi-echo spin echo sequences for T2-mapping (T2 MESE) were performed. PDw fs sequences covered the entire knee joint in coronal orientation, while the single-slice MESE sequence was positioned to align with one of the central PDw fs slices. **Table 1** lists the scan parameters.

## Arthroscopy

Meniscectomy was realized by standard arthroscopy, performed by S. N. (a common trunk-trained orthopedic surgeon), between the two imaging sessions (**Figure 1(c)**). Each knee specimen was accessed through standard anteromedial and anterolateral portals using a 4-mm 30° arthroscope. Then, a systematic diagnostic arthroscopy was conducted, and all compartments were inspected and probed with particular attention to the medial meniscus. To open the medial femorotibial compartment, manual valgus stress was applied. A medial meniscectomy was carried out segmentally from posterior to anterior using standard arthroscopic straight-tip scissors and a meniscal punch (Arthrex, Naples, FL, USA). Resections were performed radially, and the remaining fibers were assessed with an arthroscopic probe to confirm loss of load-bearing integrity. Irregular margins were trimmed with a motorized shaver to achieve a smooth rim without residual weight-bearing tissue. After the procedure, the joint was thoroughly irrigated, excess fluid was removed, and portals were closed with simple sutures. Before the post-arthroscopy imaging session, the joint was dried externally and repositioned in the loading device.

## Image analysis

### Segmentation

T.L. (medical imaging scientist with eight years of experience in medical imaging) performed the segmentations using ITK-SNAP (v. 3.6, http://www.itksnap.org/ [17]). Per joint and for all pressure and meniscectomy configurations, the two medial cartilage plates (i.e., medial tibia [MT] and medial femur [MF]) were manually segmented based on the TE=37.25 ms images of the respective T2 MESE image series (**Figure 2a**). The PDw fs image of the corresponding slice was consulted during segmentation to correctly distinguish cartilage from synovial fluid. M. v. d. S. (radiologist with 7 years of experience in clinical imaging) validated all segmentations.

### $T_2$ mapping

Per pressure state, meniscectomy state and medial segmented region, $T_2$ parametric maps were calculated in Python (v3.9.9, https://www.python.org/) using the T2 MESE image series and the respective segmentation (i.e., MT or MF) as an input. For every voxel within the segmentation mask, the measured MR signal evolution as a function of echo time, $M_{xy}(TE)$, was extracted and fitted with an exponential decay function of the form

$M_{xy}(TE) = M_0 \cdot \exp(-TE/T_2)$, [Eq.1]

where $M_0$ is a multiplicative factor, TE is the echo time and $T_2$ is the $T_2$ relaxation time. Per fitted signal decay, $r^2$ was calculated to assess the fit quality. Per segmented region, a mean T2 value and standard deviation (SD) were calculated. Voxels with unexpectedly high T2 values for cartilage (T2 > 100 ms) or with a bad fit quality ($r^2<0.8$) were discarded from this calculation.

### Radiomic features

PyRadiomics (v3.0.1, https://pyradiomics.readthedocs.io/en/latest/) was used to calculate the gray level co-occurrence matrix (GLCM) features variance (*'SumSquares'* in pyRadiomics), homogeneity (*'InverseDifference'*), contrast (*'Contrast'*) and energy (*'JointEnergy'*), in line with previous studies of our group [18]. The GLCM permits to analyze the texture of an input image within a segmented area after discretizing the image into discrete bins. In brief, its *(i,j)*-th element represents the number of occasions that two pixels with intensity *i* and *j* are neighboring each other. Variance and homogeneity relate to the local homogeneity of an image, where a high variance indicates a broad distribution of neighboring intensity level pairs about the mean intensity level in the GLCM and high homogeneity relates to more uniform gray levels. Contrast relates to local intensity variations, where high values indicate greater differences in intensity values between neighboring voxels. Energy relates to the occurrence of homogeneous patterns, where with values indicate that more instances of the same intensity value pairs neighbor each other more frequently [19]. GLCM features were separately calculated for the two segmented regions, selecting a bin width of 5 ms.

### Joint space width

Per pressure state and meniscectomy state, the medial joint space width (JSW) within the segmented slice was calculated based on the respective medial (MT and MF) segmentations to assess the success of pressure application. First, the individual extent of the two segmented regions and, in result, their overlap along the mediolateral direction were determined. The medial JSW was then measured in the center of the overlapping region as the distance along a vertical line from the most distal tibial segmented voxel to the most proximal femoral segmented voxel (**Figure 2b**).

## Statistical analysis

All statistical evaluations were performed in R (v4.3.1, https://www.R-project.org [20]). The effects of meniscectomy state (*men_state*), i.e., NT or PM, and loading configuration (*press_state*), i.e., UL, LN, or LV, on the parameters of interest were analyzed by fitting a separate linear mixed effects model ('lmer' from the R package 'lmer4') per parameter (T2, variance, homogeneity, contrast or energy) and region (MT or MF) as follows:

*Model(parameter, region) <- lmer(parameter ~ men_state + press_state + (1|specimen), data=data(parameter ,region))* [Eq. 2]

Here, the fixed effects *men_state* and *press_stat*e were nominal variables with 2 and 3 factors, respectively, while the nominal variable specimen was treated as a random variable. The dependent variable parameter was continuous in all cases. Additionally, the same data were fitted using the above-defined models with an interaction term (i.e., *men_state*press_state*) to investigate if there were any significant interaction effects. For none of the parameters and regions under investigation, significant interaction effects were found, which is why the initial models as defined by Eq. 2 were considered appropriate. Consequently, pair-wise post-hoc comparisons were performed only for the main effects, i.e., the average effect of each factor independently of the other, yielding four pairwise comparisons. Hence, we applied a four-fold Bonferroni correction to the p-values to account for the multiple comparison problem.

Similarly, a linear mixed effects model without interaction effects was fitted for the medial and lateral JSWs, Again, pair-wise post-hoc comparisons were obtained using four-fold Bonferroni correction.

Throughout all statistical evaluations, a significance level of 0.05 was selected.

# Results

## Study cohort

One knee joint had to be excluded because of complete cartilage denudation in the medial compartment. In total, 5 knee joints were measured pre and post meniscectomy as per the defined study protocol. They originated from 3 male and 2 female donors with a mean age of 80 $\pm$ 11 (SD) years and an age range 63 – 95 years.

## MRI reveals successful load application between the different configurations

**Figure 3** shows the segmented T2 maps of the MT and MF in all pressure and meniscectomy states for an example knee joint as overlays with the respective anatomical image (i.e., the TE=37.5 ms image from the MESE series). As can be seen from the anatomical images, LV visually resulted in a tightening of the joint space on the medial side and an opening on the lateral side. Between UL and LN, the expected narrowing of the medial joint space in both compartments is visually less obvious. After meniscectomy, the MF and MT compartment show a shift along the mediolateral direction with respect to each other that was not present in the intact configuration.

## Medial JSW indicates most important pressure application under varus loading and post meniscectomy

The mean medial JSW however, indicates a narrowing of the medial joint space from the UL over the LN to the LV pressure state and from the INT to the PM meniscectomy state (**Table 2, Figure 4**), that was significant for all pair-wise comparisons (UL vs. LN: $p = 0.003$, UL vs. LV: $p < 0.001$, LN vs. LV: $p < 0.001$; INT vs. PM: $p = 0.03$).

## T2 relaxation time not suitable to assess loading and meniscectomy state

When zooming into the segmented T2 maps, overall higher T2-values in for the UL-INT, UL-PM and LN-PM configurations as well as a busier textural appearance of the tibial cartilage plate become apparent (**Supplementary Figure 1**). Averaged over all knee joints, the mean T2 relaxation times and their SD (**Table 2, Figure 5(a)** [MT] and **Figure 6(a)** [MF]), however, do not reflect these visual impressions. Accordingly, linear mixed effects modelling did not reveal any statistically significant differences in the T2 relaxation times between the different pressure states and meniscectomy states, respectively, neither for the MT or the MF cartilage (**Table 4**).

## Radiomic parameters, especially variance, susceptible to pressure-induced changes in the femoral cartilage

The radiomic parameters, in contrast, showed a dependency on the pressure state, especially in the MF cartilage. Here, variance and energy (**Figure 6(b, e)**) showed a clear decrease and increase, respectively, from the UL over the LN to the LV pressure state, which proved to be significant for all pair-wise comparisons between pressure states ('contrast') and for all but UL vs. LN ('energy'), see **Table 3**. In the MT, no clear trends were visible for any of the radiomic parameters (**Figure 5(b)-(e)**), in line with mostly insignificant differences in the pair-wise comparisons (**Table 3**).

## No significant radiomic parameter changes between meniscectomy states

None of the radiomic parameters was susceptible to the meniscectomy state, neither in the MT or the MF, as per statistically non-significant p-values (all $p>0.05$, **Table 3**). With concurrent trends in the MT and MF, T2 values and variance showed a decrease with meniscectomy, while energy tended to increase with meniscectomy (**Supplementary Figures 2 and 3**). Homogeneity showed opposing trends between MT and MF, which were increasing and decreasing when going from the INT to the PM state, respectively.

## Discussion

This study assessed the effect of mechanical loading and meniscectomy on T2 mean values, four radiomic texture features of cartilage and the medial JSW in the medial compartment of the knee joint. As expected, medial JSW indicated the most important compression under varus loading (i.e., in the LV configuration) and a joint space narrowing after meniscectomy. While T2 values did not show any significant dependency on the pressure or the meniscectomy state (all pair-wise comparisons $p>0.05$), the radiomic GLCM-parameter 'variance' revealed itself particularly promising to assess loading-induced textural changes in the T2-maps of the femoral cartilage (UL vs. LN: $p=0.042$; UL vs. LV: $p<0.001$; LN vs. LV: $p=0.022$), and, in part, in the tibial cartilage (LN vs. LV: $p<0.013$). Meniscectomy did not significantly alter the T2 mean values or radiomic parameters, respectively.

In contrast to our initial expectations, T2 mean values obtained during stress MRI were not sensitive to the loading and meniscectomy condition. As far as T2 mapping under loading is concerned, the literature evidence is somewhat diverse, but holds evidence for a T2 reduction under pressure application [21, 22], especially within the medial compartment, that is mainly related to fluid reduction within the cartilage. Of note, some other previous studies exist for which T2 changes under loading were not significant, while this was the case for other compositional MR parameters such as T1rho [23, 24]. When pressure application was surrogated by different types of physical activities (e.g., simulated standing in the scanner, running, knee bends, etc.), moderate (0-5%) T2 reductions were noted, as summarized by Coburn et al. in a recent review [25], although there are even reports about increased T2 relaxation times after pronounced physical activity [26]. This suggests an advantage of controlled mechanical loading over physical activity studies, for which the activity level is more difficult to standardize.

Importantly, this study found an added value of T2-based radiomic features for the assessment of cartilage response to loading, as compared to "simple" T2 mapping only. For instance, the analyzed radiomic parameters variance, homogeneity and energy showed the expected gradual behavior from the UL to LN to LV configuration, whereas this was not the case for the mean T2 values. This is in line with a study of Janáčová et al., who found significant changes of several GLCM-based radiomic parameters, but not of the T2 values when comparing between knee cartilage lesions and reference lesions [27]. Based on T2-based radiomic features, machine learning models have been constructed to discriminate OA subjects from healthy controls [28], to predict the progression of symptomatic early OA [29], or to identify patients predisposed for posttraumatic OA after anterior cruciate ligament reconstruction [30]. In the latter study, Xie at al report an improved discriminatory performance of the radiomic model over cartilage T2 values alone and best performance when combining radiomic features and T2 values.

One important idea behind this study was to translate the results of Shiomi et al. [16], which were obtained in a porcine cadaveric knee joint model, but with a similar study design and loading device, to a human cadaveric knee joint model. The authors equally performed T2 mapping in three loading conditions and before and after meniscectomy. Yet, they acquired T2 maps for several adjacent sagittal slices that included anterior, central, and posterior cartilage regions, while the present study measured a single coronal slice in the central weight-bearing area, a choice made for reasons of overall scan time. When comparing our medial femoral T2 values to the ones obtained by Shiomi et al. in their medial central femoral compartments (termed 'M2' and 'M5'), it is striking that significant T2 decreases after load application were found in the porcine model (averaged between M2 and M5, -12.7% and -17.5% T2 decrease from UL to LN and from UL to LV before meniscectomy respectively, and, less important but still present, -9.5% and -6.8% T2 decrease after meniscectomy), but not in the present study. One explanation could root in the generally rather advanced the age of body donors vs. the probably young age of the animals.

In contrast to the device employed by Shiomi et al., our device implemented a hydraulic pressure control, which can be precisely controlled and actuated from outside the MR room over a large range of pressures [11] instead of by actioning a screw directly at the device. Besides being electronically controllable, the hydraulic pressure adjustment also facilitates the workflow, as there is no need to move the patient (or specimen) in and out of the scanner in between different loading and unloading configuration.

Cadaveric joint models constitute a unique environment to perform imaging free of movement artifacts. Also, they allow for controlled chirurgical interventions, which is generally not ethically permitted in healthy individuals. Still, the translation of findings to the in vivo setting must be the ultimate goal. An in vivo study employing stress MRI before and after meniscectomy, however, is challenging, because it could be confounded by the presence of prior knee injury that may have yielded cartilage damage prior to the enrollment of patients scheduled for meniscal surgery. Also, stress MRI is not widely available, and few commercial devices exist (e.g. Telos Stress Device, Type SE-MR, Telos GmbH, Wölfersheim, Germany). Still, comparing the effects of different physical activities between patients with and without previous meniscectomy and monitoring symptoms of OA over time may give further indications about the potential utility of stress MRI post meniscectomy in vivo. For example, Lindner et al. performed T2 mapping before and immediately after running and found significantly decreased T2 values immediately after running in patients who underwent medial partial meniscectomy, but not in healthy knees [15].

This study has limitations. First, the sample size with N=5 specimens was relatively small. This was due to the relatively high experimental workload to collect this rich dataset per specimen encompassing several loading and meniscectomy states. Second, we selected only four radiomic features for our analysis, based on previous literature. Future research may consider other texture features beyond contrast, variance, homogeneity, and energy. Third, we did not assess whether potential inter-reader differences introduced any variability into the mean T2 value or texture features. Yet, high segmentation quality was assured by having an experienced radiologist verify all segmentations. Fourth, the MRI sequence was a single slice (2D) acquisition and did therefore not permit the reconstruction of volumetric T2 data. By the time of the study, this choice was made for reasons of total scan time per specimen in relation to scanner availability; however, future studies should consider volumetric, faster T2 mapping methods with near-isotropic spatial resolution such as quantitative double echo steady state [31] or Multi-Interleaved X-prepared TSE with inTUitive Relaxometry [32, 33].

## Conclusion

T2-based radiomics showed an improvement over simple T2-mapping in assessing the response-to-loading patterns of cartilage by stress MRI under different loading configurations. This has potential implications when employing stress MRI to diagnose early OA-mediated degenerative cartilage changes in post-meniscectomy patients.

# Tables

## Table 1: Scan parameters

***Table 1**: **Acquisition parameters of the MR sequences.** Abbreviations: PDw – proton density-weighted, fs – fat saturated, MESE – multi-echo spin echo, MS – multi-slice (interleaved acquisition of multiple 2D slices), TSE – turbo spin echo, SPAIR – spectral attenuated inversion recovery, N/A – not applicable, 2D – two-dimensional, SENSE – sensitivity encoding (a parallel imaging technique)*

| Sequence | PDw fs | T2 MESE |
|---|---|---|
| Technique | TSE | TSE |
| Acquisition type | MS | 2D |
| Fat saturation | SPAIR | none |
| Orientation | coronal | coronal |
| Field of view [mm²] | 180 x 180 | 180 x 180 |
| Acquisition matrix | 368 x 364 | 368 x 368 |
| Reconstruction matrix | 720 x 720 | 720 x 720 |
| Number of slices | 30 | 1 |
| Slice thickness / slice gap [mm] | 3.0 / 0.3 | 3 / N/A |
| Scan percentage [%] | 100 | 100 |
| Repetition time [ms] | 5643 | 1400 |
| Echo time [ms] | 30 | $n$ * 7.45 |
| Echo train length | N/A | 15 |
| Flip angle [°] | 90 | 90 |
| TSE factor | 13 | 15 |
| Bandwidth [Hz / pixel] | 238 | 273 |
| Number of signal averages | 1 | 1 |
| Scan time [min:s] | 05:27 | 08:38 |

## Table 2: Descriptive statistics

***Table 2**: **Descriptive statistics results.** Mean values and standard deviations of the quantitative T2 relaxation time, the GLCM-based radiomic parameters variance, homogeneity, contrast, and energy, respectively (indicated for the medial tibial and medial femoral region), and the medial JSW for the different pressure and meniscectomy states. Values are indicated as mean value ± SD. Abbreviations: INT – intact (prior to meniscectomy), JSW – joint space width, LN – neutral loading, LV – varus loading, MC – medial compartment, MF – medial femur, MT – medial tibia, PM – post meniscectomy, UL - unloaded*

| Parameter | Region | Meniscectomy state | Pressure state | | |
|---|---|---|---|---|---|
| | | | UL | LN | LV |
| T2 [ms] | MT | INT | 31 ± 5 | 32 ± 3 | 31 ± 3 |
| | | PM | 30 ± 4 | 32 ± 6 | 30 ± 4 |
| | MF | INT | 33 ± 1 | 35 ± 6 | 33 ± 4 |
| | | PM | 31 ± 2 | 34 ± 5 | 32 ± 4 |
| Variance | MT | INT | 7 ± 1 | 8 ± 2 | 5 ± 1 |
| | | PM | 6 ± 2 | 6 ± 3 | 5.2 ± 0.9 |
| | MF | INT | 14 ± 2 | 11 ± 4 | 8 ± 3 |
| | | PM | 14 ± 3 | 10 ± 3 | 7 ± 2 |
| Homogeneity | MT | INT | 0.66 ± 0.04 | 0.64 ± 0.04 | 0.64 ± 0.04 |
| | | PM | 0.67 ± 0.03 | 0.64 ± 0.06 | 0.66 ± 0.03 |
| | MF | INT | 0.56 ± 0.03 | 0.54 ± 0.02 | 0.58 ± 0.02 |
| | | PM | 0.54 ± 0.03 | 0.54 ± 0.02 | 0.57 ± 0.02 |
| Contrast | MT | INT | 2.0 ± 0.4 | 2.3 ± 0.6 | 1.9 ± 0.4 |
| | | PM | 1.7 ± 0.3 | 1.9 ± 0.7 | 1.6 ± 0.3 |
| | MF | INT | 3.1 ± 0.8 | 3.3 ± 0.5 | 2.7 ± 0.4 |
| | | PM | 3.4 ± 0.7 | 3.4 ± 0.4 | 2.6 ± 0.6 |
| Energy | MT | INT | 0.04 ± 0.02 | 0.037 ± 0.007 | 0.04 ± 0.01 |
| | | PM | 0.05 ± 0.02 | 0.04 ± 0.02 | 0.05 ± 0.01 |
| | MF | INT | 0.018 ± 0.003 | 0.019 ± 0.004 | 0.024 ± 0.007 |
| | | PM | 0.019 ± 0.003 | 0.020 ± 0.002 | 0.027 ± 0.005 |
| Medial JSW [mm] | MC | INT | 4.4 ± 0.6 | 3.3 ± 0.6 | 3.3 ± 0.4 |
| | | PM | 3.6 ± 0.5 | 2.5 ± 0.4 | 2. 5 ± 0.3 |

## Table 3: Statistical analysis results (JSW)

***Table 3: Pair-wise post hoc test results quantifying the influence of pressure state and meniscectomy state on the joint space width in the medial compartment.*** *The table indicates the p-values after 4-fold Bonferroni-correction. Statistically significant p-values are indicated in* ***bold text****. Abbreviations: LN – neutral loading, LV – varus loading, men_state – meniscectomy state, MF – medial tibia, MT – medial tibia, press_state – pressure state, UL – unloaded.*

| Averaged effect | Comparison | p-value |
|---|---|---|
| men_state | UL - LN | **0.003** |
| | UL - LV | **<0.001** |
| | LN - LV | **<0.001** |
| press_state | INT - PM | **0.030** |

## Table 4: Statistical analysis results (T2/Radiomics)

***Table 4: Pair-wise post hoc test results quantifying the influence of pressure state and meniscectomy state on T2 relaxation times and radiomic parameters in the regions MT and MF.*** *The table indicates the p-values after 4-fold Bonferroni-correction. Statistically significant p-values are indicated in* ***bold text.***

*Abbreviations: LN – neutral loading, LV – varus loading, men_state – meniscectomy state, MF – medial tibia, MT – medial tibia, press_state – pressure state, UL – unloaded.*

| Region | Averaged effect | Comparison | T2 [ms] | Variance | Homogeneity | Contrast | Energy |
|---|---|---|---|---|---|---|---|
| MT | men_state | UL - LN | 1.000 | 1.000 | 0.139 | 0.449 | 0.243 |
| | | UL - LV | 1.000 | 0.168 | 1.000 | 1.000 | 1.000 |
| | | LN - LV | 1.000 | **0.013** | 1.000 | 0.139 | 0.850 |
| | press_state | INT - PM | 1.000 | 0.476 | 0.474 | 0.138 | 0.113 |
| MF | men_state | UL - LN | 0.720 | **0.042** | 1.000 | 1.000 | 1.000 |
| | | UL - LV | 1.000 | **P<0.001** | **0.035** | 0.060 | **0.001** |
| | | LN - LV | 0.549 | **0.022** | **0.002** | **0.043** | **0.006** |
| | press_state | INT - PM | 1.000 | 1.000 | 0.594 | 1.000 | 1.000 |

# Figures

## Figure 1: Experimental setup

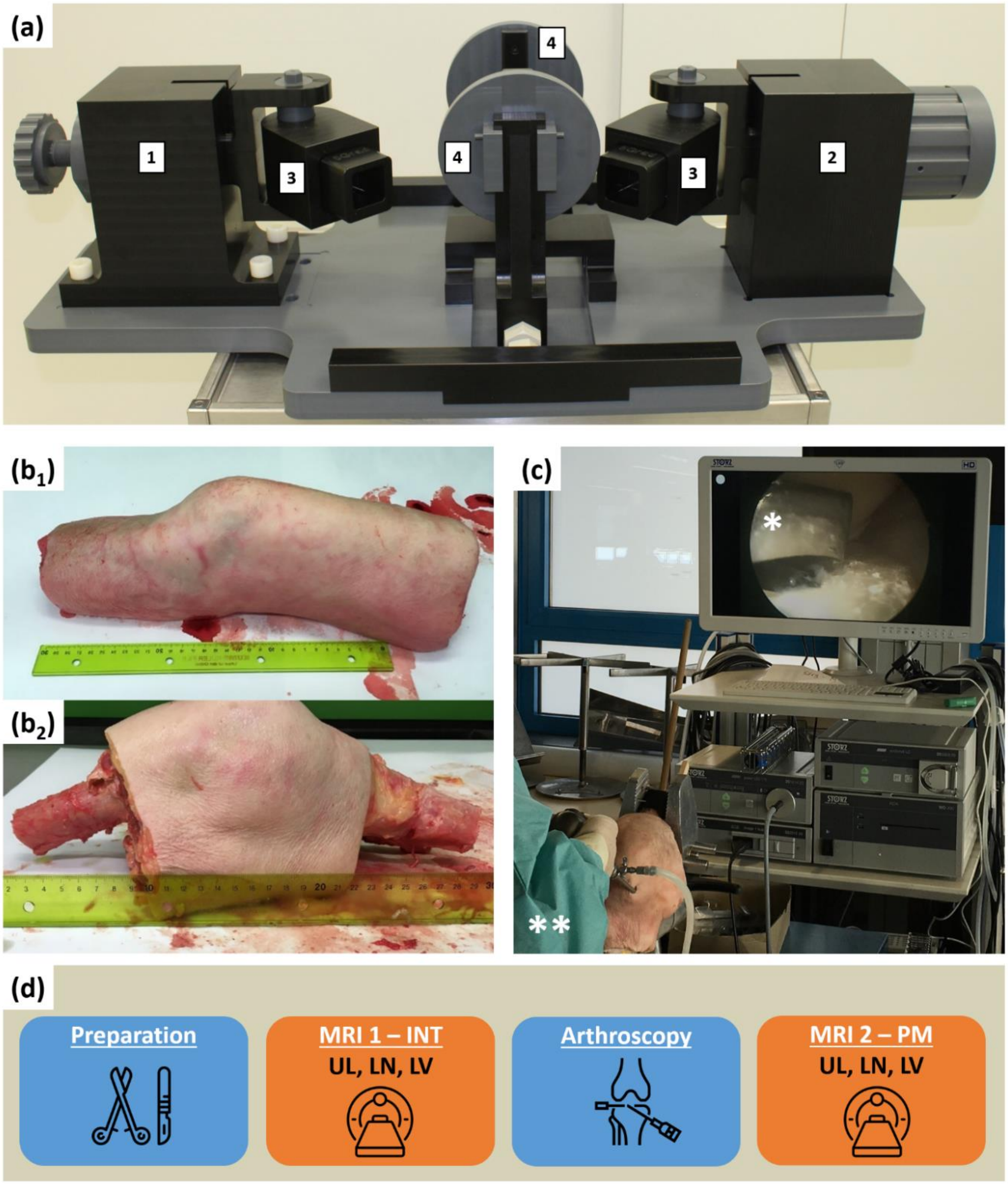

**Figure 1** (a) Loading apparatus. 1: Rigidly fixed femoral side with an adjusting screw to account for various specimen lengths, 2: tibial side with pneumatic pressure control, 3: rotatable mounting blocks with bone pots to mount the femoral and tibial bone, respectively, that can be fixed at different angulations, 4: varus-valgus positioning unit with press-point disks that hold the specimen in place under load application. (b) Specimen preparation. (b1) shows an intact, defrozen knee joint, (b2) shows the same knee joint after preparation for mounting into the bone pots. (c) Meniscectomy by means of arthroscopy. *: meniscus punch, **: arthroscopist. (d) Workflow. After specimen preparation, the first set of MRI measurements was performed with the intact knee joint (INT) in the three loading configurations (UL, LN, LV). After arthroscopy, the MRI measurements were performed a second time with the knee joint in the post meniscectomy state (PM).

## Figure 2: Image processing methods

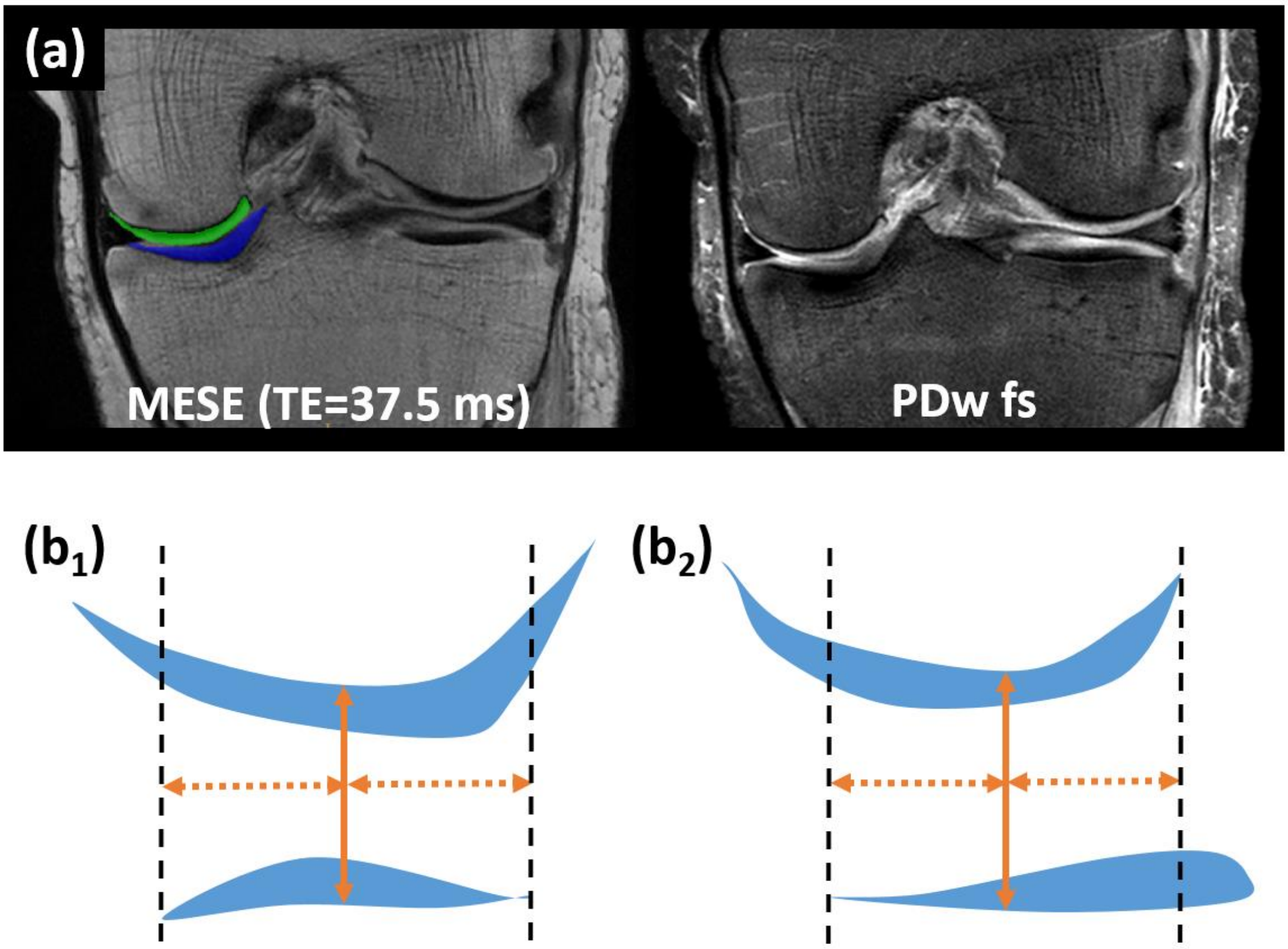

**Figure 2: Image processing methods.** (a) Example segmentation, showing an overlay of the segmented femoral (green region) and tibial (blue region) medial cartilage plates with the 5$^{th}$ echo time image (TE = 37.5 ms) of the MESE series. The PDw fs image was additionally consulted to distinguish cartilage from synovial fluid. (b) JSW calculation for two example geometrical situations. ($b_1$): the femoral cartilage plate extends over the tibial cartilage plate to both sides, ($b_2$): both cartilage plates are shifted with respect to each other. In both cases, the two black dashed lines indicate the overlap of the medial femoral and tibial cartilage plate along the mediolateral direction. The medial JSW (indicated by the solid orange arrow) was calculated as the vertical distance of the most distal tibial voxel and the most proximal femoral voxel as measured in the center of the mediolateral overlapping zone (indicated by orange dotted arrows).

## Figure 3: T2 mapping

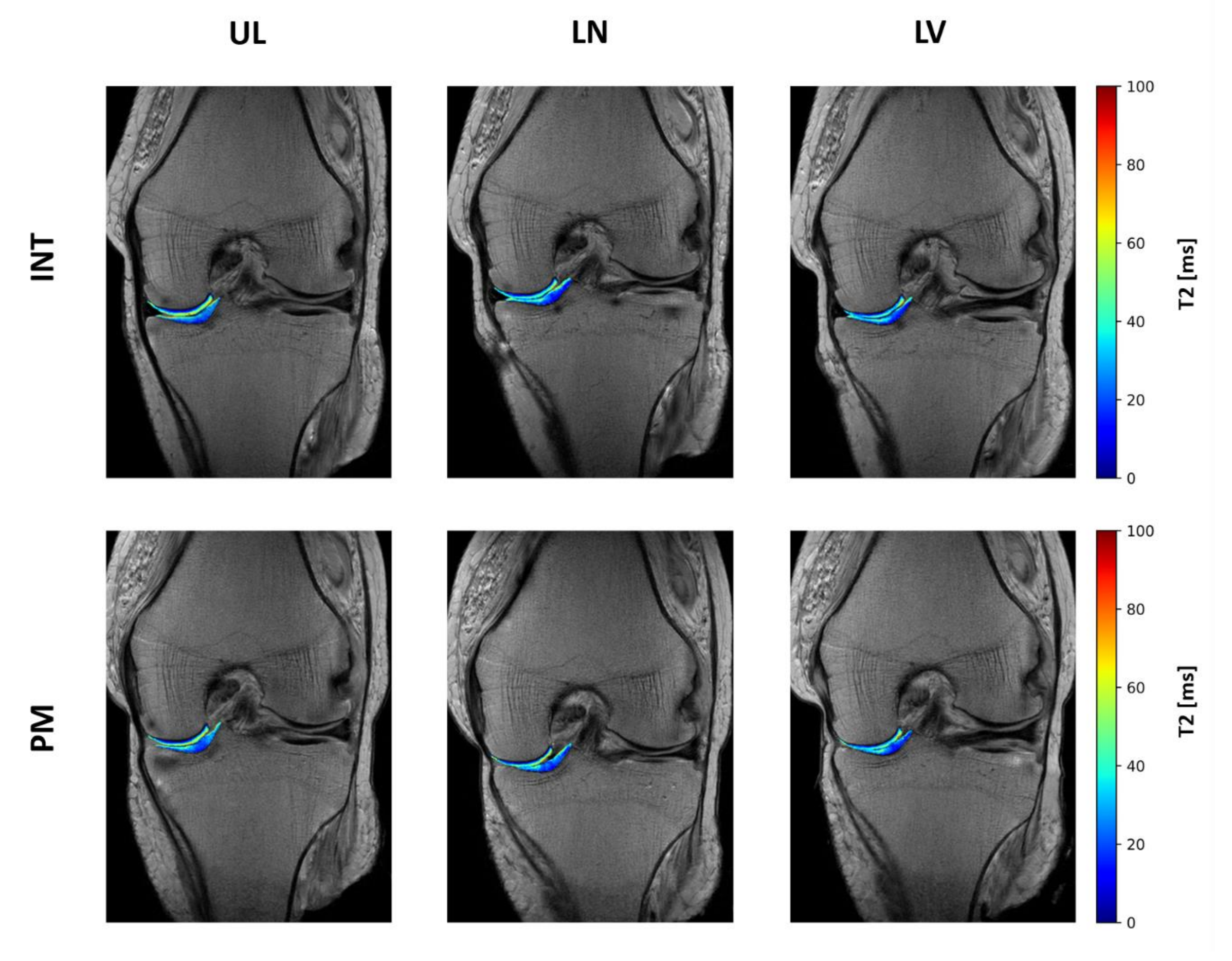

***Figure 3***: *Overlays of the T2 maps in the two segmented regions (MT and MF) and the 5$^{th}$ echo time image (TE=37.5 ms) of the MESE image series for an example knee joint. From left to right, columns depict the three investigated pressure states (UL, LN and LV), while rows from top to bottom correspond to the two meniscectomy states (INT and PM).*

*Abbreviations: MF – medial femur, MT – medial tibia, INT – intact meniscus (before meniscectomy), PM – post meniscectomy, LN – loading under neutral axis alignment, LV – loading under varus axis alignment, UL – unloaded.*

## Figure 4: Barplots JSW

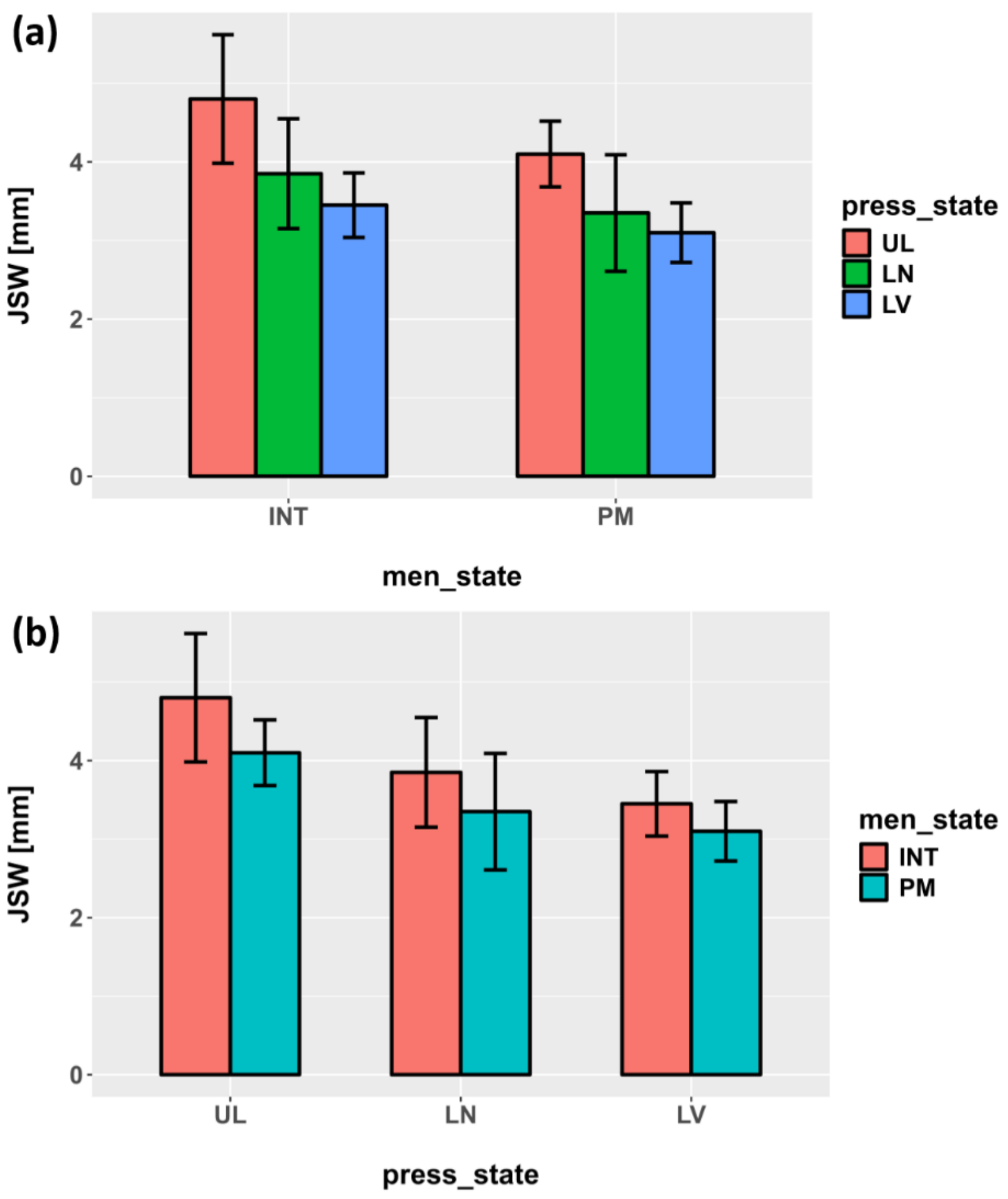

**Figure 4: Influence of pressure state and meniscectomy state on the medial JSW**. (a) and (b) show the mean medial JSW and their SD, grouped by the meniscectomy state or the pressure state, respectively.

Abbreviations: men_state – meniscectomy state, press_state – pressure state, JSW – joint space width, SD – standard deviation.

## Figure 5: Barplots T2/Radiomics (MT)

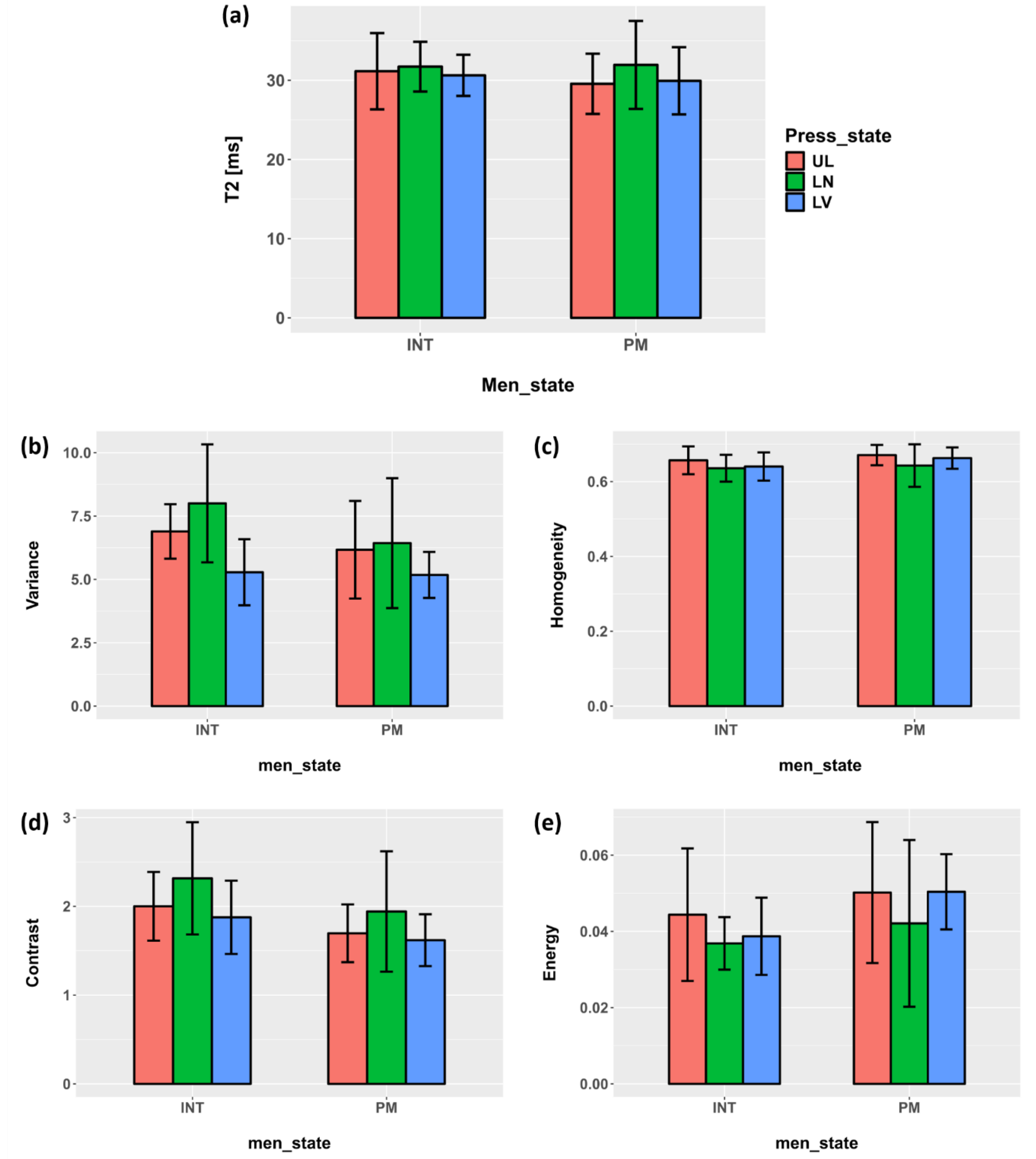

***Figure 5: Influence of meniscectomy and pressure state on (a) T2-values and (b-e) the radiomic parameters variance, homogeneity, contrast and energy in the medial tibia.*** *Bars and error bars depict mean value and SD, respectively.*

*Abbreviations: INT – intact, LN – neutral loading, LV – varus loading, men_state – meniscectomy state, MF – medial femur, MT – medial tibia, PM – post meniscectomy, press_state – pressure state, SD – standard deviation, UL - unloaded.*

# Figure 6: Barplots T2/Radiomics (MF)

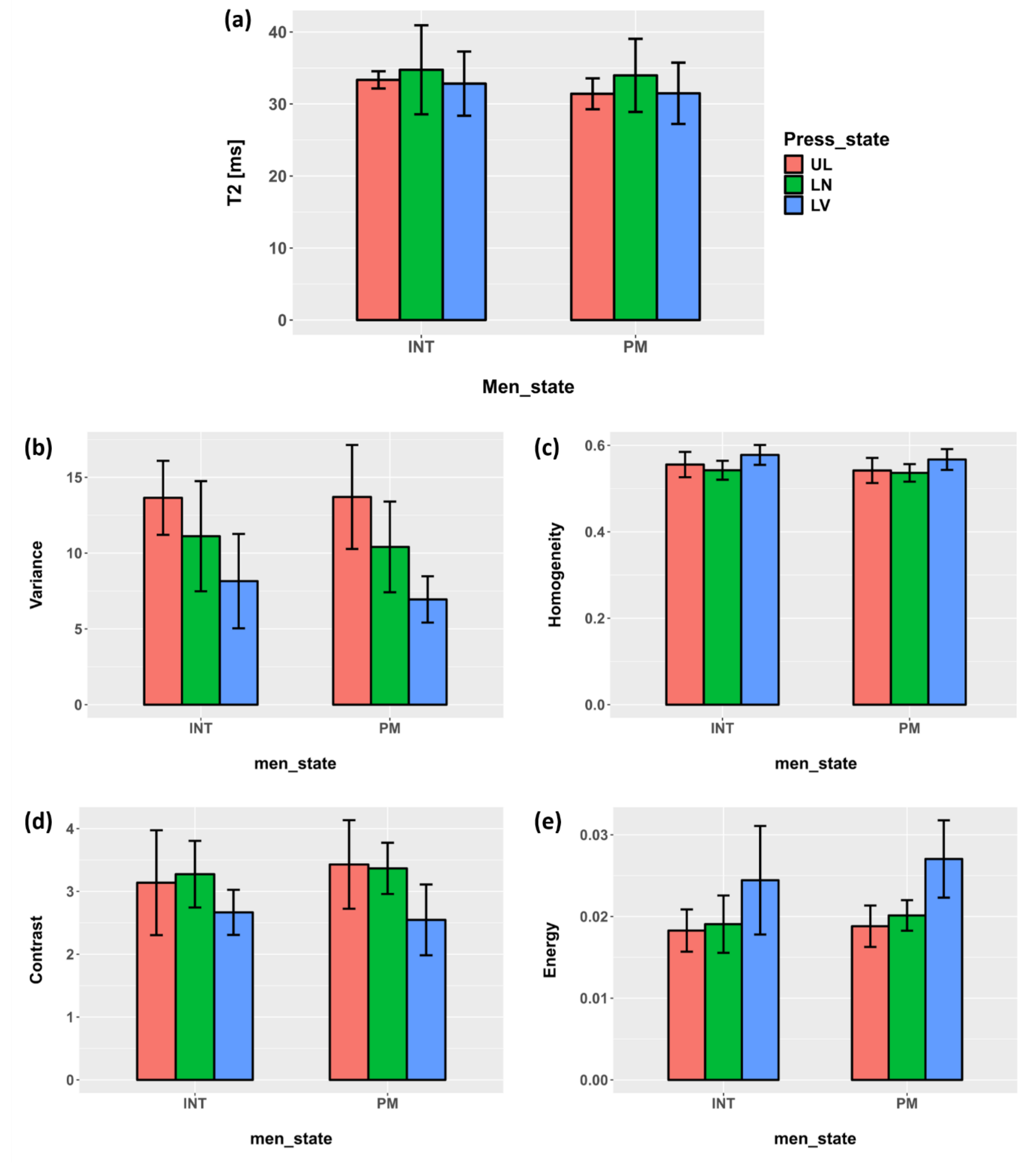

***Figure 6: Influence of meniscectomy and pressure state on (a) T2-values and (b-e) the radiomic parameters variance, homogeneity, contrast and energy in the medial femur.*** *Bars and error bars depict mean value and SD, respectively.*

*Abbreviations: INT – intact, LN – neutral loading, LV – varus loading, men_state – meniscectomy state, MF – medial femur, MT – medial tibia, PM – post meniscectomy, press_state – pressure state, SD – standard deviation, UL - unloaded.*

# Supplementary Material

## Supplementary Figure 1: Close-up T2-Mapping

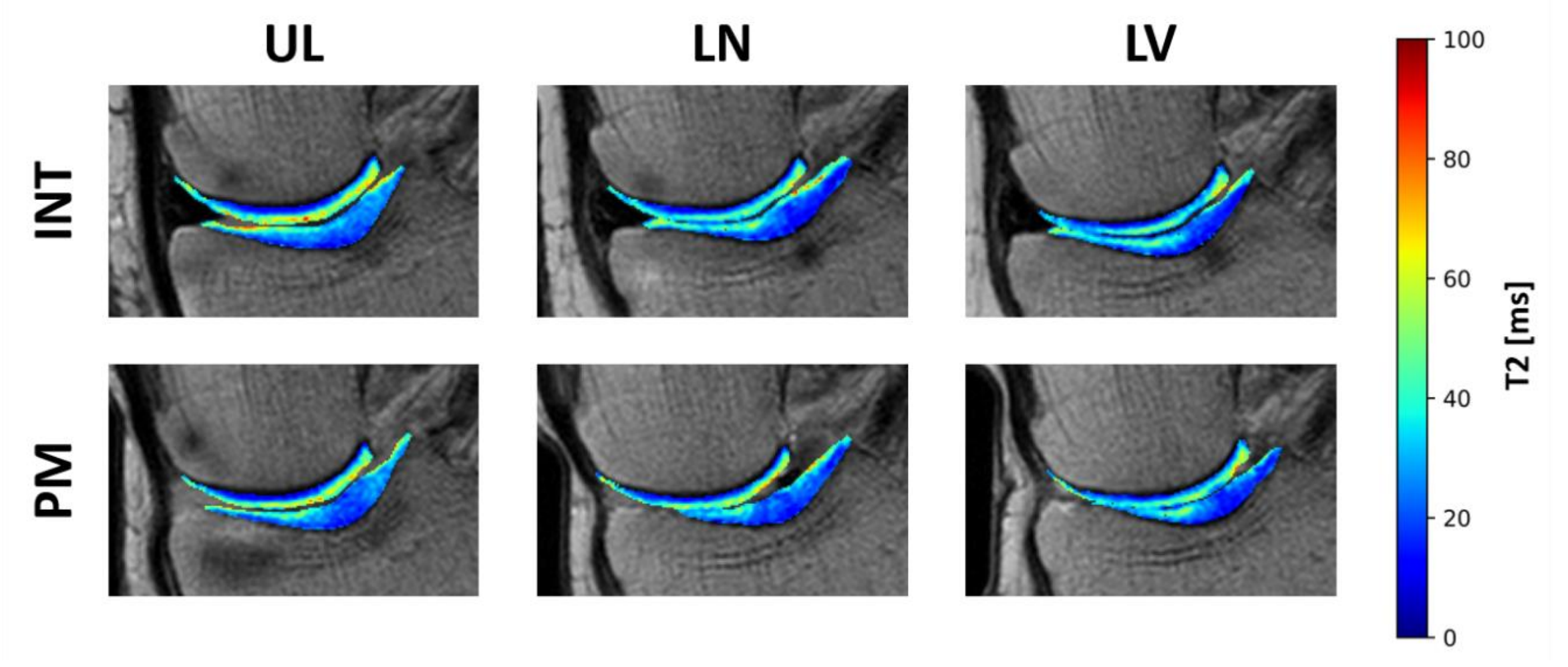

***Supplementary Figure 1****: Close-up on the medial joint space area for the same example knee joint that is presented in* ***Figure 3****. For further explanations, please see* ***Figure 3****.*

## Supplementary Figure 2: Bar plots sorted by pressure state (MT)

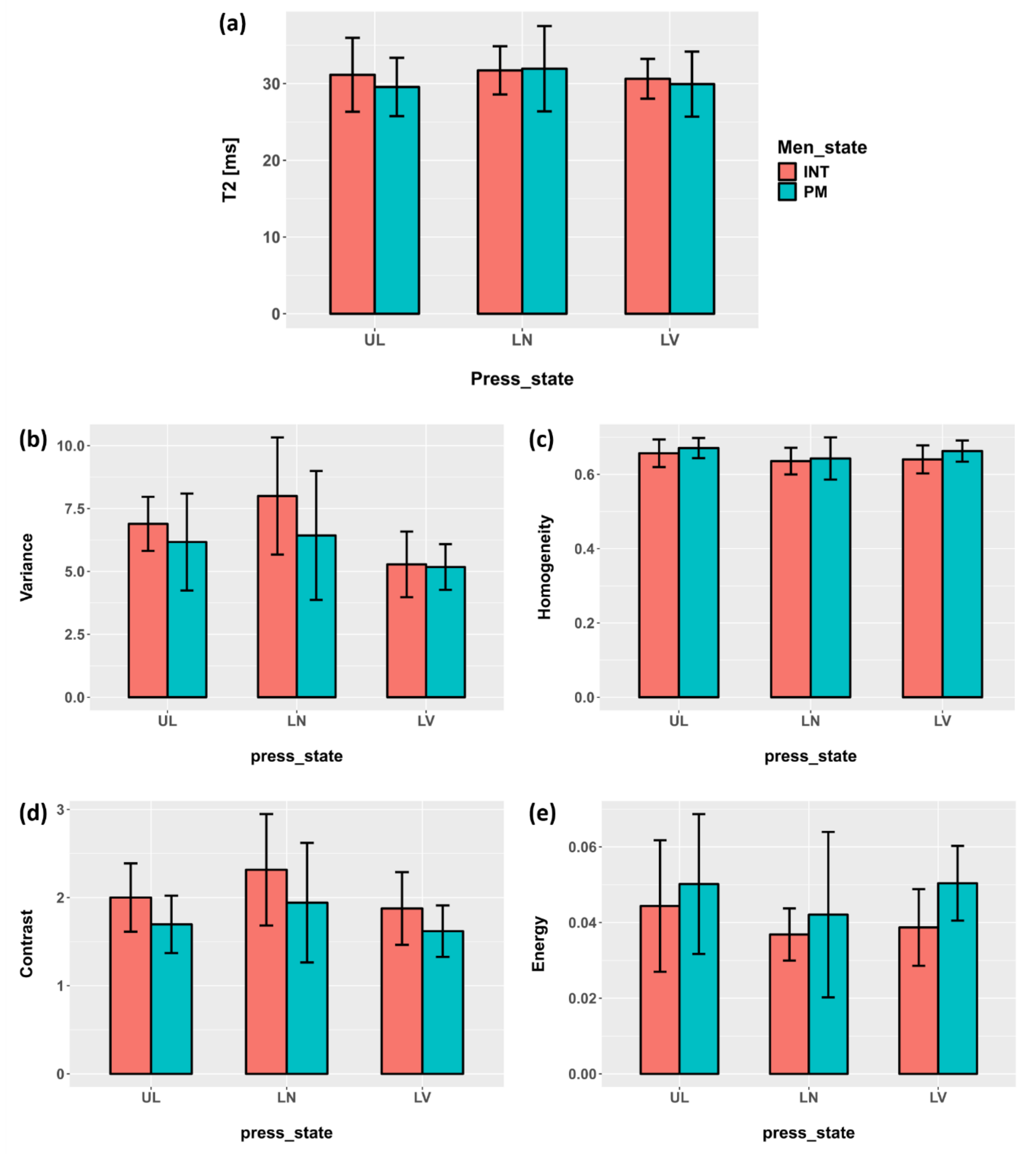

***Supplementary Figure 2: Bar plots for same data as shown in Figure 5, but sorted by press_state.***

## Supplementary Figure 3: Bar plots sorted by pressure state (MF)

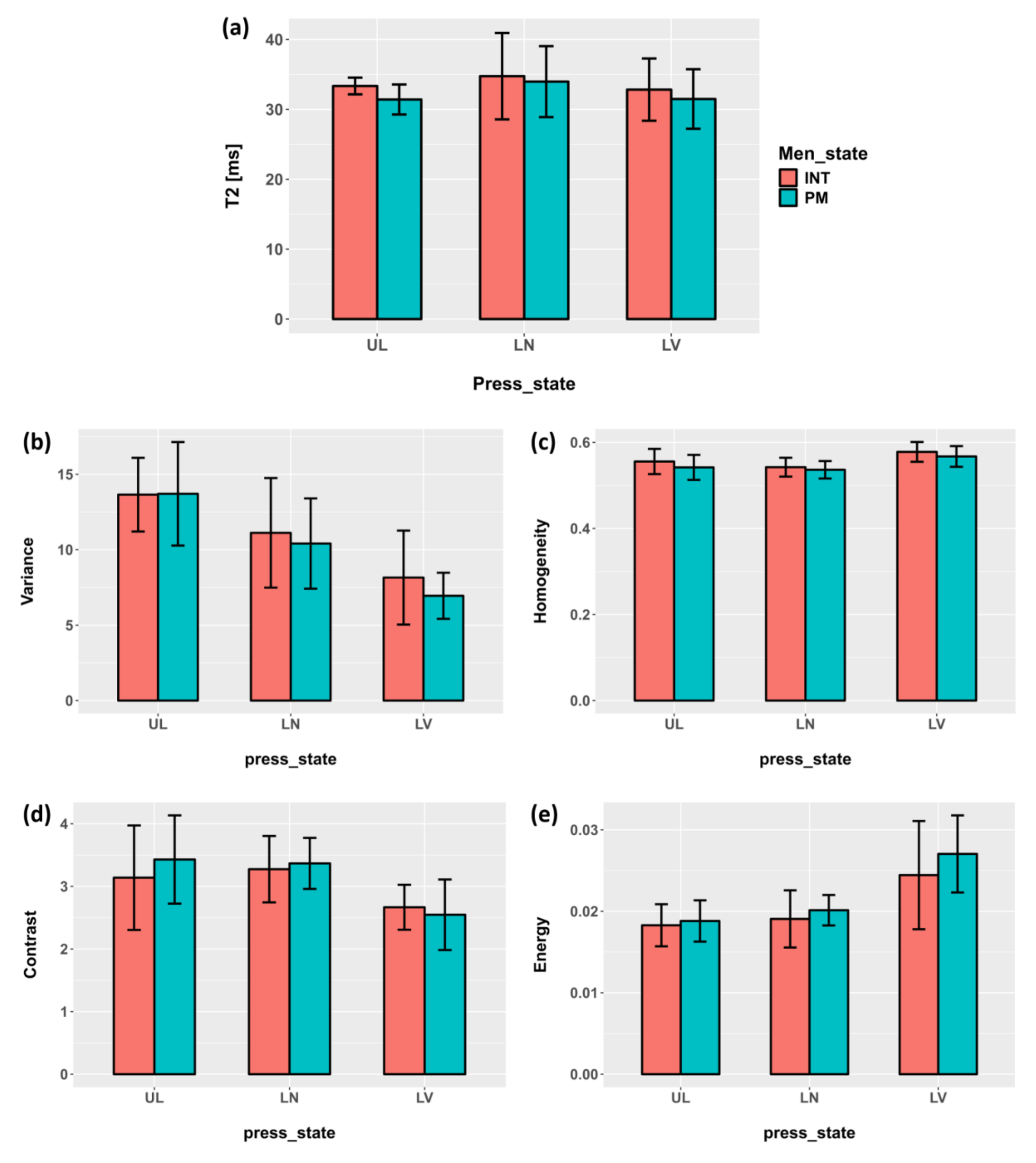

***Supplementary Figure 3: Bar plots for same data as shown in Figure 6, but sorted by press_state.***